\documentclass[aps,prl,twocolumn,superscriptaddress,showpacs,preprintnumbers,amsmath,amssymb]{revtex4}

\usepackage{graphicx}
\usepackage{dcolumn} 
\usepackage{color}   

\graphicspath{{ps}}

\begin{document}

%\preprint{\vbox{ \hbox{   }
%	\hbox{Belle Preprint 2007-XX}
%	\hbox{KEK   Preprint 2007-XX}
%}}

\title{ \quad\\[1.0cm] Search for the $CP$-violating decays 
       $\Upsilon(4S) \to B^0\bar{B}^0 \to J/\psi K^0_S + J/\psi(\eta_c)K^0_S$}

% authorlist
\affiliation{Budker Institute of Nuclear Physics, Novosibirsk}
\affiliation{Chiba University, Chiba}
\affiliation{University of Cincinnati, Cincinnati, Ohio 45221}
\affiliation{Department of Physics, Fu Jen Catholic University, Taipei}
\affiliation{Justus-Liebig-Universit\"at Gie\ss{}en, Gie\ss{}en}
\affiliation{The Graduate University for Advanced Studies, Hayama}
\affiliation{Hanyang University, Seoul}
\affiliation{University of Hawaii, Honolulu, Hawaii 96822}
\affiliation{High Energy Accelerator Research Organization (KEK), Tsukuba}
\affiliation{Hiroshima Institute of Technology, Hiroshima}
\affiliation{Institute of High Energy Physics, Chinese Academy of Sciences, Beijing}
\affiliation{Institute of High Energy Physics, Vienna}
\affiliation{Institute of High Energy Physics, Protvino}
\affiliation{Institute for Theoretical and Experimental Physics, Moscow}
\affiliation{J. Stefan Institute, Ljubljana}
\affiliation{Kanagawa University, Yokohama}
\affiliation{Korea University, Seoul}
\affiliation{Kyungpook National University, Taegu}
\affiliation{Ecole Polyt\'ecnique F\'ed\'erale Lausanne, EPFL, Lausanne}
\affiliation{University of Ljubljana, Ljubljana}
\affiliation{University of Maribor, Maribor}
\affiliation{University of Melbourne, School of Physics, Victoria 3010}
\affiliation{Nagoya University, Nagoya}
\affiliation{Nara Women's University, Nara}
\affiliation{National Central University, Chung-li}
\affiliation{National United University, Miao Li}
\affiliation{Department of Physics, National Taiwan University, Taipei}
\affiliation{H. Niewodniczanski Institute of Nuclear Physics, Krakow}
\affiliation{Nippon Dental University, Niigata}
\affiliation{Niigata University, Niigata}
\affiliation{University of Nova Gorica, Nova Gorica}
\affiliation{Osaka City University, Osaka}
\affiliation{Osaka University, Osaka}
\affiliation{Panjab University, Chandigarh}
\affiliation{Princeton University, Princeton, New Jersey 08544}
\affiliation{Saga University, Saga}
\affiliation{University of Science and Technology of China, Hefei}
\affiliation{Seoul National University, Seoul}
\affiliation{Sungkyunkwan University, Suwon}
\affiliation{University of Sydney, Sydney, New South Wales}
\affiliation{Tata Institute of Fundamental Research, Mumbai}
\affiliation{Toho University, Funabashi}
\affiliation{Tohoku Gakuin University, Tagajo}
\affiliation{Tohoku University, Sendai}
\affiliation{Department of Physics, University of Tokyo, Tokyo}
\affiliation{Tokyo Institute of Technology, Tokyo}
\affiliation{Tokyo Metropolitan University, Tokyo}
\affiliation{Tokyo University of Agriculture and Technology, Tokyo}
\affiliation{Virginia Polytechnic Institute and State University, Blacksburg, Virginia 24061}
\affiliation{Yonsei University, Seoul}
  \author{O.~Tajima}\affiliation{High Energy Accelerator Research Organization (KEK), Tsukuba} % KEK
  \author{M.~Hazumi}\affiliation{High Energy Accelerator Research Organization (KEK), Tsukuba} % KEK
  \author{I.~Adachi}\affiliation{High Energy Accelerator Research Organization (KEK), Tsukuba} % KEK
  \author{H.~Aihara}\affiliation{Department of Physics, University of Tokyo, Tokyo} % Tokyo
  \author{V.~Aulchenko}\affiliation{Budker Institute of Nuclear Physics, Novosibirsk} % BINP
  \author{T.~Aushev}\affiliation{Ecole Polyt\'ecnique F\'ed\'erale Lausanne, EPFL, Lausanne}\affiliation{Institute for Theoretical and Experimental Physics, Moscow} % ITEP
  \author{A.~M.~Bakich}\affiliation{University of Sydney, Sydney, New South Wales} % Sydney
  \author{E.~Barberio}\affiliation{University of Melbourne, School of Physics, Victoria 3010} % Melbourne
  \author{A.~Bay}\affiliation{Ecole Polyt\'ecnique F\'ed\'erale Lausanne, EPFL, Lausanne} % Lausanne
  \author{I.~Bedny}\affiliation{Budker Institute of Nuclear Physics, Novosibirsk} % BINP
% ...
% Ljubljana
\collaboration{The Belle Collaboration}
\noaffiliation
%% end author list

\begin{abstract}

We report the first search for
$CP$ violating decays of the $\Upsilon(4S)$
using a data sample that contains 535 million
$\Upsilon(4S)$ mesons
with the Belle detector at the KEKB asymmetric-energy $e^+ e^-$
collider.
A partial reconstruction technique is employed to
enhance the signal sensitivity.
No significant signals were observed.
We obtain an upper limit of 
$ 
4
\times 10^{-7} $
 at the 90~\% confidence level
for the
branching fractions of 
the $CP$ violating modes,
$
 \Upsilon(4S) 
 \to B^0\bar{B}^0
 \to 
 J/\psi K^0_S + J/\psi(\eta_c) K^0_S
$.
Extrapolating the result, we find that 
an observation with 5$\sigma$ significance is expected
with a 
30
ab$^{-1}$ data sample, which is within the reach of
a future super $B$ factory.

\end{abstract}

\pacs{11.30.Er, 12.15.Hh, 13.25.Gv, 13.25.Hw}

\maketitle

\tighten

{\renewcommand{\thefootnote}{\fnsymbol{footnote}}}
\setcounter{footnote}{0}

$CP$ violation has been established 
in the neutral kaon system \cite{ref_cpv_kaon}
and the neutral $B$ meson system \cite{ref_sin2phi1}.
In the standard model (SM)
Kobayashi-Maskawa theory,
it arises from an irreducible phase in the weak interaction 
quark-mixing matrix \cite{KM}.
This theory predicts that 
$CP$ violation in the $\Upsilon(4S)$ system should also exist.

In the decay 
$
\Upsilon(4S) \to B^0 \bar{B}^0 \to f_1 f_2
$,
where $f_1$ and $f_2$ are $CP$ eigenstates,
the $CP$ eigenvalue of the final state
$f_1 f_2$ is $\xi=-\xi_1\xi_2$.
Here the minus sign corresponds to odd parity
from the angular momentum between $f_1$ and $f_2$.
If $f_1$ and $f_2$ have the same $CP$ eigenvalue,
i.e. 
$
(\xi_1, \xi_2)
~=~(+1, +1)
$
or $(-1, -1)$, $\xi$ is equal to $-1$.
Such decays,
for example 
$
(f_1, f_2) = (J/\psi K^0_S, J/\psi K^0_S)
$,
violate $CP$ conservation since the
$\Upsilon(4S)$ meson has
$J^{PC} = 1^{--}$ and thus has
$\xi_{\Upsilon(4S)} = +1$.
The branching fraction within the SM is
\begin{eqnarray}
&&
{\cal B}(~\Upsilon(4S) \to B^0\bar{B}^0 \to f_1 f_2~) 
\nonumber
\\
&& 
\hspace{-12pt}
 =
  F \cdot
  {\cal B}(\Upsilon(4S) \to B^0\bar{B}^0) 
  {\cal B}(B^0 \to f_1) {\cal B}(\bar{B}^0 \to f_2)
,
\end{eqnarray}
where $F$ is a suppression factor due to $CP$ violation.
The factor $F$ can be calculated in terms of mixing and $CP$ violating parameters~\cite{bib_F},
\begin{eqnarray}
    F &\simeq& \frac{x^2}{1+x^2} (2 \sin2\phi_1)^2
     \\ &=& 0.68 \pm 0.05, \nonumber
\end{eqnarray}
where $x = \Delta m_d/\Gamma = 0.776 \pm 0.008$~\cite{bib_PDG}, 
$\Delta m_d$ is the $B^0$ mixing parameter,
$\Gamma$ is the average decay width of the neutral $B$ meson.
The angle $\phi_1$ is one of the three interior angles of the unitarity
triangle of the quark-mixing matrix,
and $\sin2\phi_1 = 0.675 \pm 0.026$~\cite{bib_PDG}.
The effect of direct $CP$ violation is neglected in this formula.
The same expression also holds 
for the case in which $f_1$ and 
$f_2$ are different final states both of which
are governed by
$b \to c\bar{c}s$ transitions; examples include
$\eta_c K^0_S$, $\psi(2S) K^0_S$ and $\chi_{c1} K^0_S$.

In this Letter, 
we present the first search for 
$CP$ violating decays of the $\Upsilon(4S)$.
The data sample used 
contains 535 million $\Upsilon(4S)$ mesons
collected  with the Belle detector at the KEKB asymmetric-energy
$e^+e^-$ (3.5 on 8~GeV) collider~\cite{KEKB}.
The Belle detector is a large-solid-angle magnetic
spectrometer that
consists of a silicon vertex detector (SVD),
a 50-layer central drift chamber (CDC), an array of
aerogel threshold Cherenkov counters (ACC), 
a barrel-like arrangement of time-of-flight
scintillation counters (TOF), and an electromagnetic calorimeter
comprised of CsI(Tl) crystals (ECL) located inside 
a superconducting solenoid coil that provides a 1.5~T
magnetic field.  An iron flux-return located outside of
the coil is instrumented to detect $K_L^0$ mesons and to identify
muons (KLM).  The detector
is described in detail elsewhere~\cite{Belle}.
Two inner detector configurations were used. A 2.0 cm radius beampipe
and a 3-layer silicon vertex detector were used for the first sample
of 152 million $B\bar{B}$ pairs, while a 1.5 cm radius beampipe, a 4-layer
silicon detector and a small-cell inner drift chamber were used to record  
the remaining 383 million $B\bar{B}$ pairs\cite{svd2}.

The identity of each charged track is determined by a sequence of
likelihood ratios that determine the hypothesis that best matches the
available information.
Tracks are identified as pions or kaons based on their specific ionization in
the CDC as well as the TOF and ACC responses.
This classification is superseded if the track is identified as a
lepton:
electrons are identified by the presence of a matching ECL cluster with
energy and transverse profile consistent with an electromagnetic shower;
muons are identified by their range and transverse scattering in the KLM.

We use $2.68 \times 10^5$
Monte Carlo (MC) simulation 
events for each signal category.
For background MC events, 
we use a sample of $3.9 \times 10^{10}$
generic $B\bar{B}$ decays in which one of
the $B$ mesons decays to a known
$J/\psi (\mu^+\mu^-~{\rm or}~e^+e^-) X$ final state.
For the dataset used in the present analysis, the MC simulation predicts 
a small signal yield,
0.04 events,
when we choose the combination
$(f_1,~f_2) = \left(J/\psi K^0_S, ~ J/\psi K^0_S\right)$
and fully reconstruct both $J/\psi K^0_S$ final states.
Here we use the
$J/\psi \to e^+e^-$, $\mu^+\mu^-$ 
and
$K^0_S \to \pi^+\pi^-$ 
modes.
In order to increase the signal yield, we instead adopt a partial reconstruction method.
We fully reconstruct one $B^0 \to J/\psi K^0_S$ decay (called $f_{J/\psi K^0_S}$ hereafter)
and find another $K^0_S$ (called $^{\rm tag}K^0_S$ hereafter) from the remaining particles.
We then reconstruct the recoil mass ($M^{\rm recoil}$) using $J/\psi K^0_S$ and $^{\rm tag}K^0_S$.
The recoil mass distribution should in principle include peaks that correspond to the
$\eta_c$, $J/\psi$, $\chi_{c1}$, or $\psi(2S)$.
We choose two of the possible combinations,
$(f_1,~f_2)$ = 
$\left(f_{J/\psi K^0_S}, ~ J/\psi ~^{\rm tag}K^0_S\right)$ 
and
$\left(f_{J/\psi K^0_S}, ~ \eta_c ~^{\rm tag}K^0_S\right)$.
In the following, 
these are referred to as
inclusive-$J/\psi$ combinations and an inclusive-$\eta_c$ combinations,
respectively. 
Based on a MC study,
we expect that the signal yield 
will increase by a factor of 
40
compared to full reconstruction 
while maintaining a reasonable
signal to background ratio (S/B) 
of about 
1/7
for these two combinations. 
We do not use other combinations because
the S/B ratio is less than 
1/100.

We use oppositely charged track pairs 
to reconstruct
$J/\psi \to e^+e^-,~\mu^+\mu^-$ decays,
where at least one track is positively identified as a lepton.
Photons within 50~mrad of the $e^+$ and $e^-$ tracks 
are included in the invariant mass calculation (denoted as $e^+e^-(\gamma)$).
The invariant mass is required to lie in the range
$-0.15 {\rm ~GeV/}c^2 < M_{ee(\gamma)} - m_{J/\psi} < 0.036 {\rm ~GeV/}c^2$
and
$-0.06 {\rm ~GeV/}c^2 < M_{\mu\mu} - m_{J/\psi} < 0.036 {\rm ~GeV/}c^2$,
where $m_{J/\psi}$ denotes the nominal mass of $J/\psi$,
$M_{ee(\gamma)}$
and
$M_{\mu\mu}$
are the reconstructed invariant masses
from $e^+e^-(\gamma)$ and $\mu^+\mu^-$, respectively.
Asymmetric intervals are used to include part of the radiative tails.
Candidate $K^0_S \to \pi^+ \pi^-$ decays are oppositely charged track pairs
that have an invariant mass within 
$\pm 0.016 {\rm ~GeV/}c^2$ ($\simeq 4 \sigma$)
of the nominal $K^0$ mass.
The $\pi^+\pi^-$ vertex is required to be displaced 
from the interaction point
in the direction of the pion pair momentum for $^{\rm tag}K^0_S$.

\begin{figure}[t]
\includegraphics[width=0.238\textwidth]{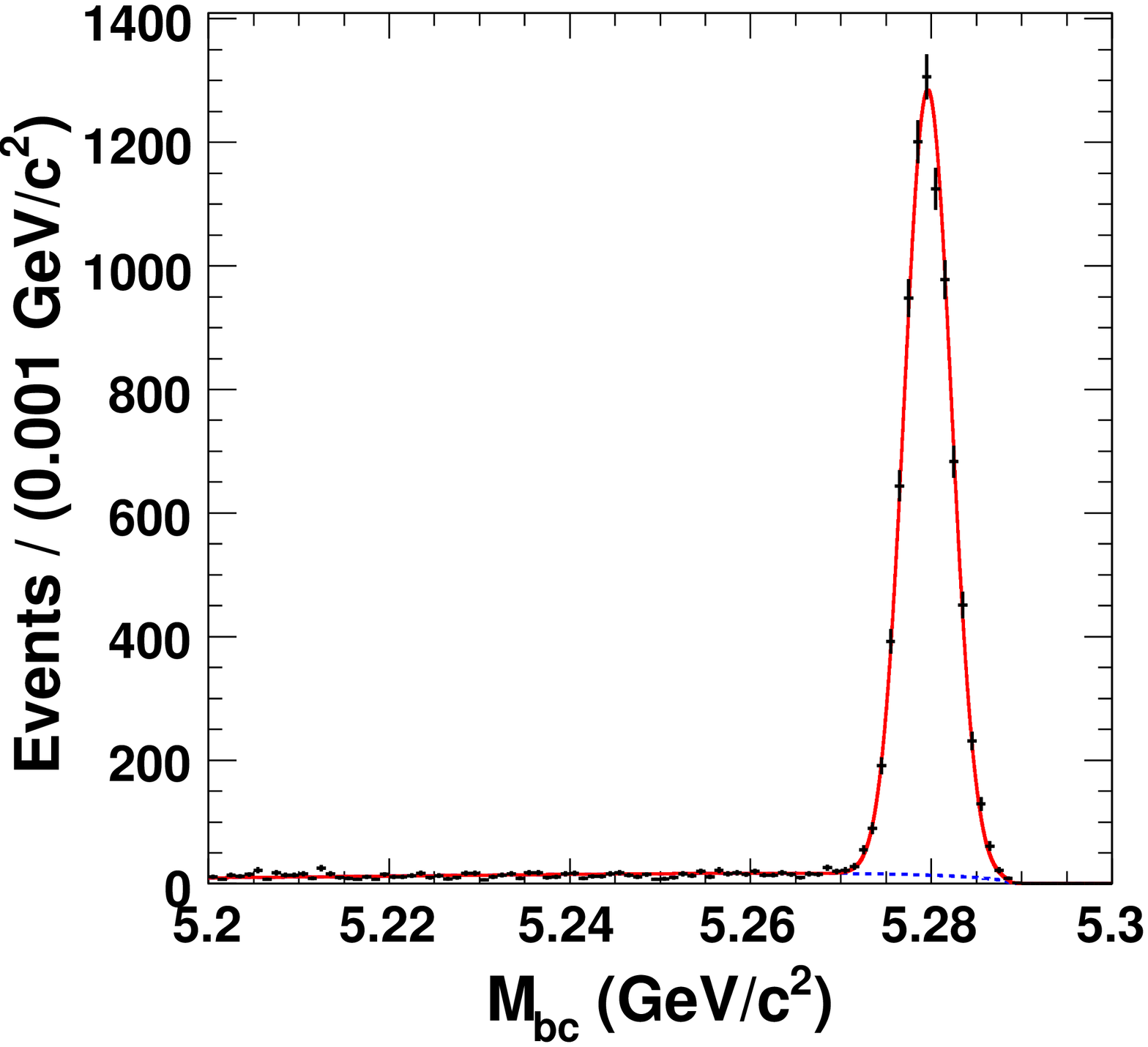}
\includegraphics[width=0.238\textwidth]{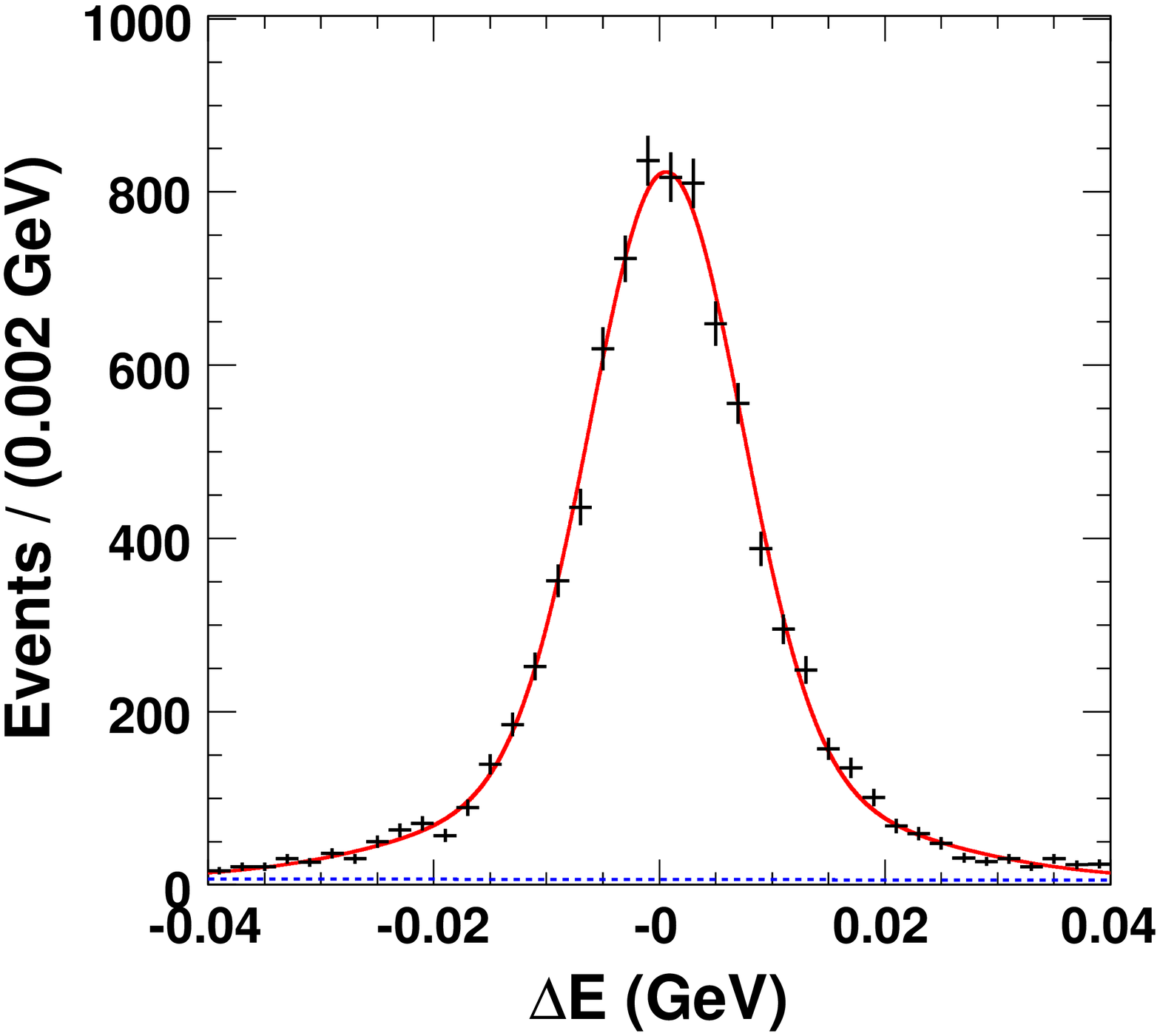}
\caption{
 $M_{\rm bc}$ (left) and $\Delta E$ (right) distributions for $B^0 \to J/\psi(\ell^+\ell^-)
 K^0_S(\pi^+\pi^-)$ decay ($l=e,\mu$).
 The solid curves show the fits to signal plus background
 distributions, and the dashed curves show the background distributions.
}
\label{fig_mbcde_jpsiks}
\end{figure}

For the full reconstruction of a $B$ decay,
we use
the energy difference 
$\Delta E \equiv E_{B}^{\rm cms} - E_{\rm beam}^{\rm cms}$
and the beam-energy constrained mass
$M_{\rm bc} \equiv \sqrt{ 
\left(
E_{\rm beam}^{\rm cms}
\right)^2
-
\left(
p_{B}^{\rm cms}
\right)^2
}
$,
where $E_{\rm beam}^{\rm cms}$ is the beam energy in the center-of-mass
system (cms) of the $\Upsilon(4S)$ resonance,
and 
$E_{B}^{\rm cms}$
and
$p_{B}^{\rm cms}$
are the cms energy and momentum of the reconstructed $B$ candidate, 
respectively.
The $M_{\rm bc}$ and $\Delta E$ distributions are shown in
Fig.~\ref{fig_mbcde_jpsiks}.
The signal 
is extracted from 
an unbinned extended maximum
likelihood fit to the $M_{\rm bc}$-$\Delta E$ distribution.
The signal shape is modeled with a single (double) Gaussian
while the background shape is modeled with an ARGUS function~\cite{bib_argus_function}
(a first order polynomial)
for the $M_{\rm bc}$ ($\Delta E$) distribution.
We obtain $8283~\pm~94$ $f_{J/\psi K^0_S}$ events
when we do not require a $^{\rm tag}K^0_S$.

We require 
$5.27 {\rm ~GeV/}c^2 \leq M_{\rm bc} \leq 5.29 {\rm ~GeV/}c^2$
and 
$| \Delta E | \leq 0.04$ GeV for $f_{J/\psi K^0_S}$.
The recoil mass is calculated
by combining 
a $f_{J/\psi K^0_S}$ candidate 
and 
a $^{\rm tag}K^0_S$ candidate.
The expected number of signal events estimated from MC
is 1.1 (0.6) with 
a reconstruction efficiency of 
28.8~(26.8)~\%
for the inclusive-$J/\psi$ $(\eta_c)$ combination
where branching fractions of sub-decays are not included.
With the partial reconstruction technique,
the number of 
$J/\psi \to e^+e^-,~\mu^+\mu^-$ decays 
in the 
($J/\psi K^0_S, ~J/\psi K^0_S$)
combination
is about twice as large as that for 
the 
($J/\psi K^0_S, ~\eta_c K^0_S$)
combination.
A total of 1.7 signal events are then expected in our dataset.

The dominant source of background is generic $B^0$ decays.
A partially reconstructed $B$ candidate should be
flavor non-specific if it is a signal event.
On the other hand, about a half of the generic $B^0$ decays that survive
the selection are flavor specific.
In order to distinguish between the signal and the background,
we therefore identify the flavor of the partially-reconstructed
accompanying $B$ meson using
leptons, charged pions and kaons
that are not associated with the fully reconstructed $B$ meson.
The procedure for flavor tagging is described in Ref.~\cite{TaggingNIM}.
We use an event-by-event flavor-tagging dilution factor, $r$, 
which ranges from $r=0$ for no flavor discrimination 
to $r=1$ for perfect flavor assignment.

We determine the signal yield by performing 
an unbinned extended maximum-likelihood fit 
to the candidate events.
The likelihood function is 
\begin{eqnarray}
 {\cal L} = \frac{1}{N!}
  \exp\left(- \sum_{k} n_{k}\right)
  \prod_{i=1}^{N}
  \left[
   {\sum_{k} n_k f_{k}(M_i^{\rm recoil}, r_i) }
  \right],
\end{eqnarray}
where 
$N$ is the total number of candidate events, 
$n_{k}$ is the number of events 
and $f_k$ is the probability density function (PDF) for each event 
category $k$, which is 
inclusive-$J/\psi$,
inclusive-$\eta_c$ or background.
The parameters 
$M_i^{\rm recoil}$ and $r_i$ are the recoil mass and $r$ value for the $i$-th
event.
The PDFs are obtained from the MC simulation.
The recoil mass distributions are modeled with 
a triple Gaussian for each signal mode
and 
an exponential shape for background.
We do not find any peaking background
in
either the MC samples 
or in
the $M_{\rm bc}$ sideband data.
The PDFs for the $r$ distributions are histograms with 10 bins
obtained from MC.
The ratio between  the inclusive-$J/\psi$ and $\eta_{c}$ signals
is fixed from the MC.

\begin{figure}[thb]
\includegraphics[width=0.28\textwidth]{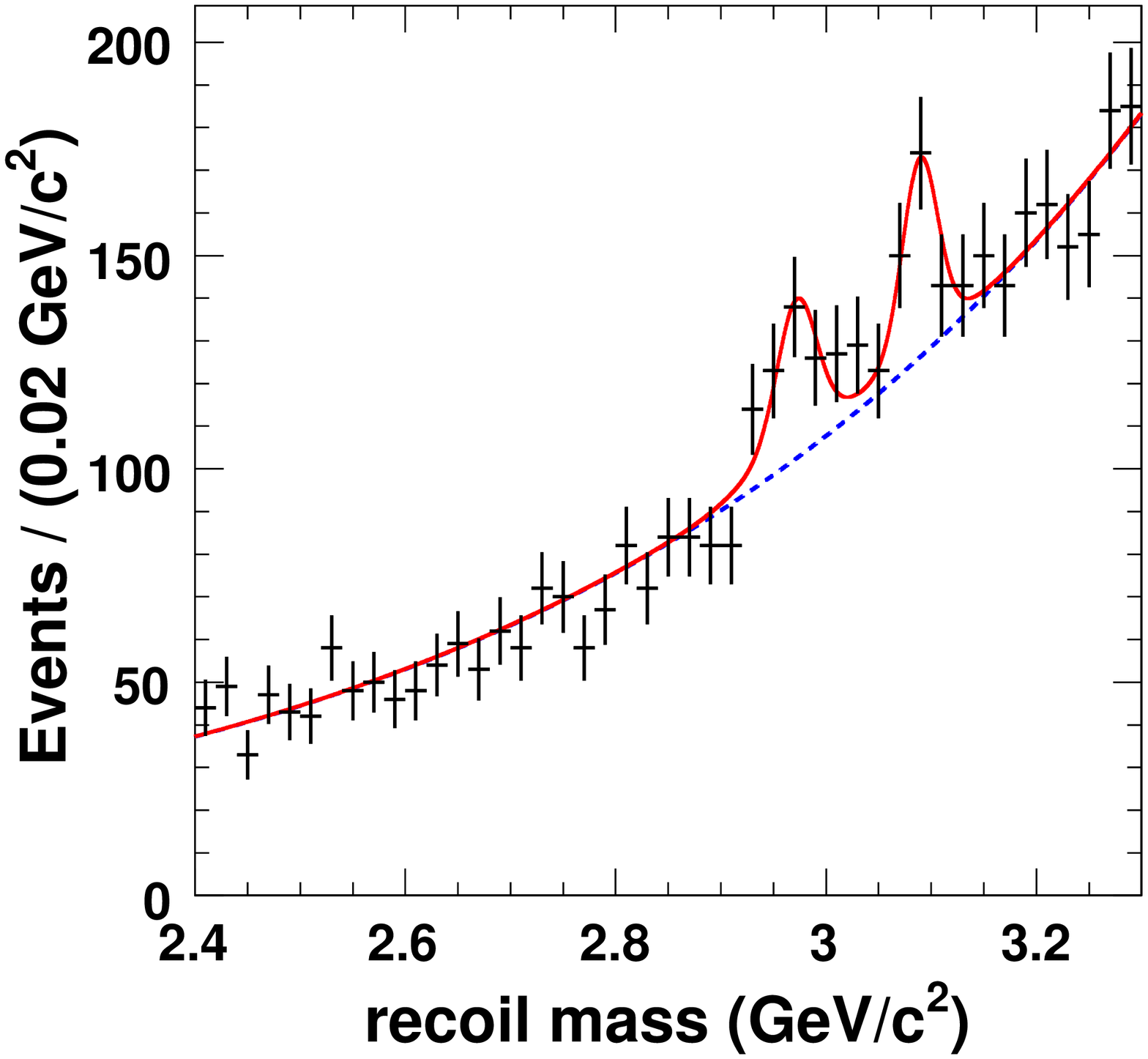}
\includegraphics[width=0.28\textwidth]{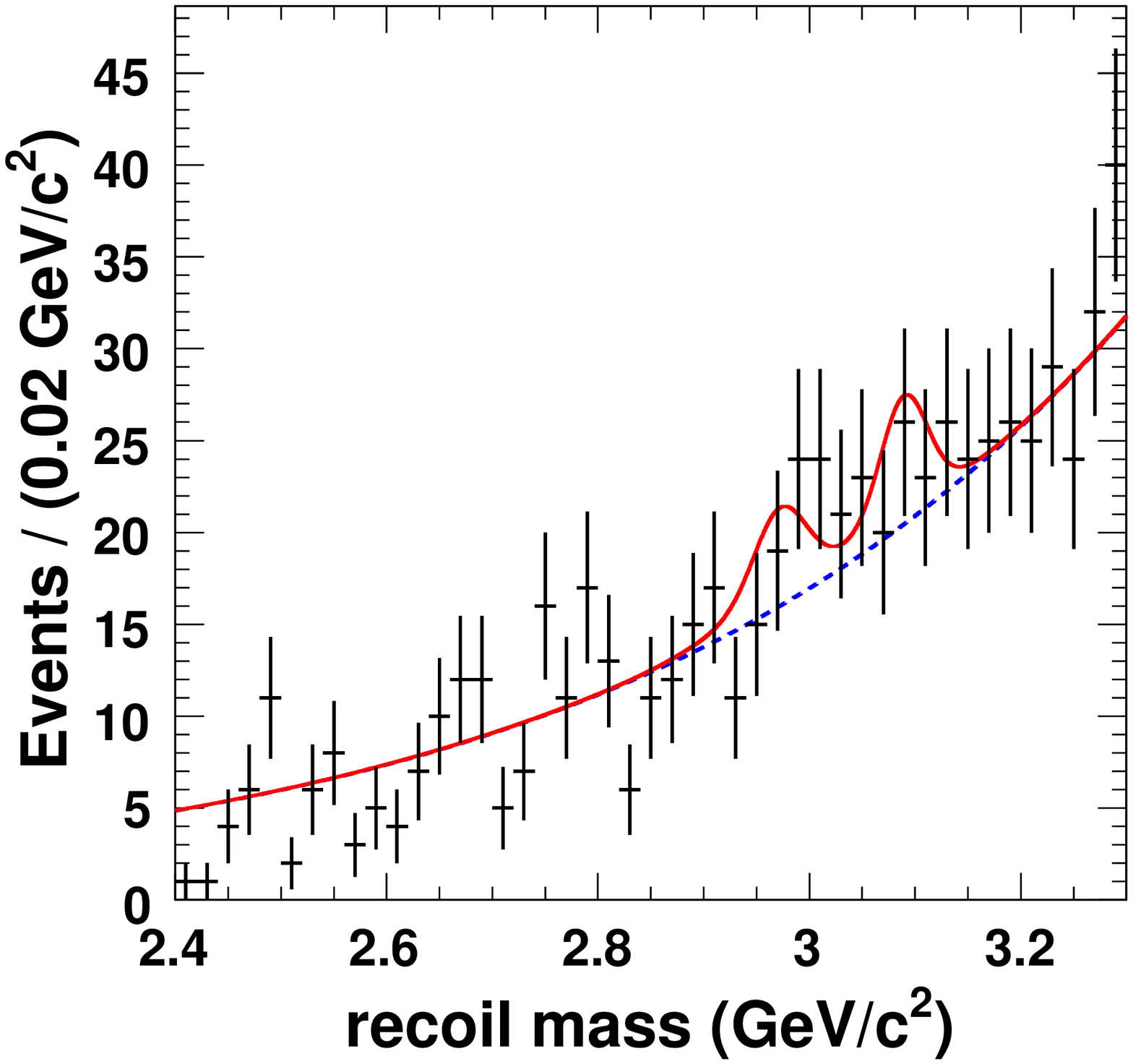}
\includegraphics[width=0.28\textwidth]{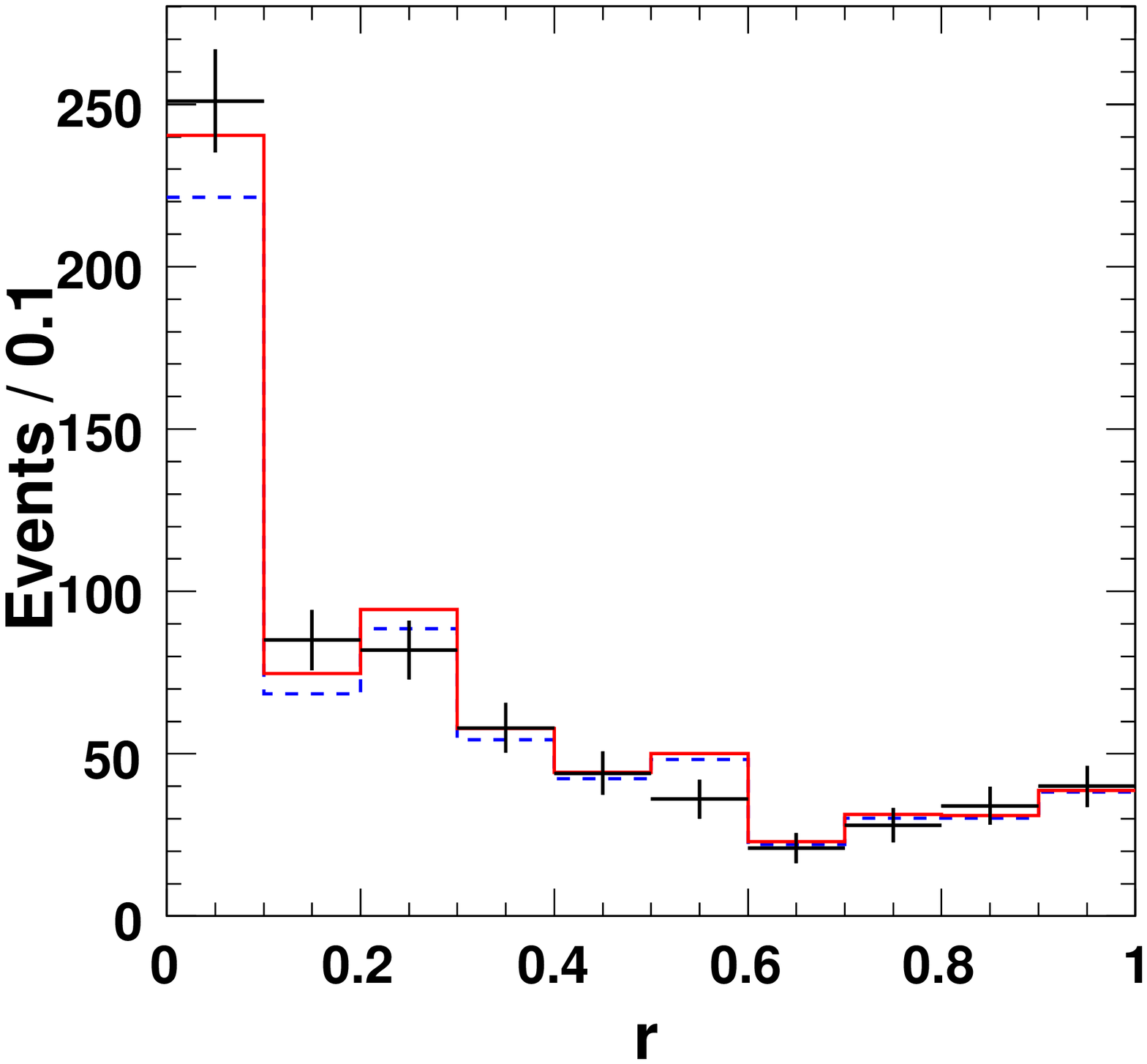}
\caption{
 Recoil mass distribution for the charged $B$ decay control samples (top),
 recoil mass (middle) and $r$ (bottom) distribution 
 for the neutral $B$ decay control samples.
 The solid curve shows the fit to signal plus background
 distributions while the dashed curve shows the background distribution.
}
 \label{fig_frecBpm_mrec_r}
\end{figure}

We check the method using 
charged $B$ decay control samples,
 $\Upsilon(4S) 
 \to 
 B^+B^-
 \to 
 \left(
 f_{B^+},~ J/\psi ~^{\rm tag}K^-
 ~{\rm and}~\eta_c ~^{\rm tag}K^-
 \right)
 $, 
 where $f_{B^+}$ 
stands for
$J/\psi(e^+e^-, \mu^+\mu^-) K^+$ and 
$\bar{D}^0(K^+\pi^-, K^+\pi^-\pi^+\pi^-)\pi^+$ decays~\cite{CC}.
Figure~\ref{fig_frecBpm_mrec_r} shows 
the recoil mass distribution for the charged $B$ control samples.
The fit yields
$206 \pm 57$ signal events, which is in good agreement with
the MC expectation (183 events).
If we float the ratio between the inclusive-$J/\psi$ and $\eta_c$ modes,
we obtain $96 \pm 23$ and $109 \pm 25$ events for the inclusive-$J/\psi$ and
$\eta_c$ modes, respectively.
These results are also consistent with the MC expectation, 90 (93) events for
inclusive-$J/\psi$ ($\eta_c$) mode.
We obtain correction factors, the mean and width for the signal peaks 
and the slope for background, by fitting these samples.

\begin{table}[b]
\caption{ Systematic uncertainties in the branching fraction measurement.}
\label{table_syst}
\begin{tabular}
{@{\hspace{0.5cm}}l@{\hspace{0.5cm}} @{\hspace{0.5cm}}r@{\hspace{0.5cm}}}
\hline \hline
Source & (\%) \\
\hline
Recoil mass distribution   & 20.5 \\
$r$ distribution           &  4.2 \\
Reconstruction efficiency  &  5.7 \\
Number of $B\bar{B}$ pairs &  1.3 \\
Branching fractions of sub-decays & 
10.9
\\
\hline
Total & 
24.3
\\
\hline \hline
\end{tabular}
\end{table}
We adopted a blind analysis method and estimated
systematic uncertainties before obtaining the final result.
The systematic uncertainties for the combined branching fraction,
$
 {\cal B}\left(
 \Upsilon(4S) 
 \to B^0\bar{B}^0
 \to 
 J/\psi K^0_S , (J/\psi, \eta_c) K^0_S
 \right),
$
are summarized in Table~\ref{table_syst}.
The dominant source of systematics is due to the
uncertainties in the correction factors for the recoil
mass distribution;
we assign 20.5~\%, which is the sum in quadrature of
19.7~\% from the signal shapes and 5.5~\% from the background shape.

Possible differences between data and the MC in the $r$ distributions
are also studied.
We use neutral $B$ decay control samples,
 $\Upsilon(4S) 
 \to 
 B^0\bar{B}^0
 \to
 \left(
 f_{B^0},~ (J/\psi, \eta_c) ~^{\rm tag}K^0_S
 \right)
 $ 
 decays,
where $f_{B^0}$ represents $B^0 \to D^{(*)-} \pi^+$ and $D^{*-} \rho^+$
followed by the
decays
$D^{*-} \to \bar{D}^0 \pi^-$,
$\bar{D}^0 \to K^+\pi^-, ~ K^+\pi^-\pi^0, ~ K^+\pi^-\pi^+\pi^-$,
$D^{-} \to K^+ \pi^- \pi^-$, 
$\rho^+ \to \pi^+ \pi^0$
and $\pi^0 \to \gamma\gamma$.
We obtain $35 \pm 16$ signal events for these samples,
which is consistent with the MC prediction
(64 events) within two standard deviations.
There is no discrepancy between data and fit results either
in recoil mass or in the $r$ distributions, as shown in Fig~\ref{fig_frecBpm_mrec_r}.
We repeat the fit using the background $r$ PDF
determined from the data in the recoil mass sideband regions
$M^{\rm recoil} \in$ (2.40, 2.85) and (3.20, 3.30) GeV/$c^2$.
The difference between the two fit results (2.6~\%) is included
in the systematic error from the $r$ distribution.
We also repeat the fit without using the $r$ distribution,
which yields a result that differs by 3.3~\% from
the nominal fit result.
We assign a 4.2~\% systematic uncertainty for the $r$ distribution,
which is the sum in quadrature of these two errors.

Systematic uncertainties from event reconstruction are 
studied by varying the particle identification, $K^0_S$ selection 
and other requirements.
The resulting changes in the signal yield
in data and MC for 
$B^0 \to J/\psi K^0_S$ 
and
$B^+ \to J/\psi K^+$
are used to estimate the systematic error.
In total, 5.7~\% of the systematic uncertainty 
that is obtained from the sum in quadrature of differences between data and MC
is assigned for event reconstruction.
The uncertainty in the total number of $B\bar{B}$ pairs is 1.3~\%.
Uncertainties in the daughter branching fractions~\cite{bib_PDG}
are dominated by those for the $\eta_c$ decays.

\begin{figure}[t]
\includegraphics[width=0.28\textwidth]{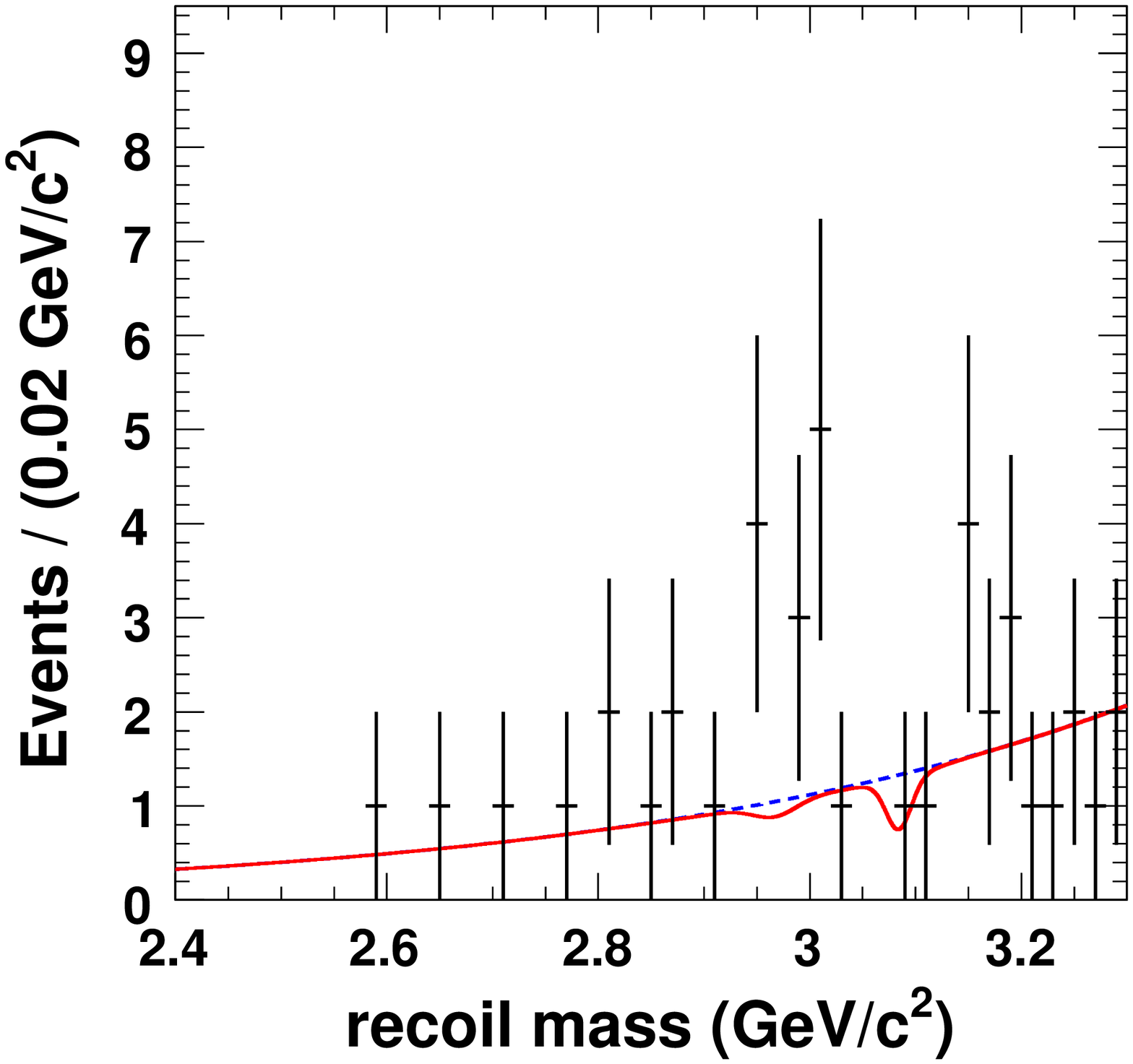}
\includegraphics[width=0.28\textwidth]{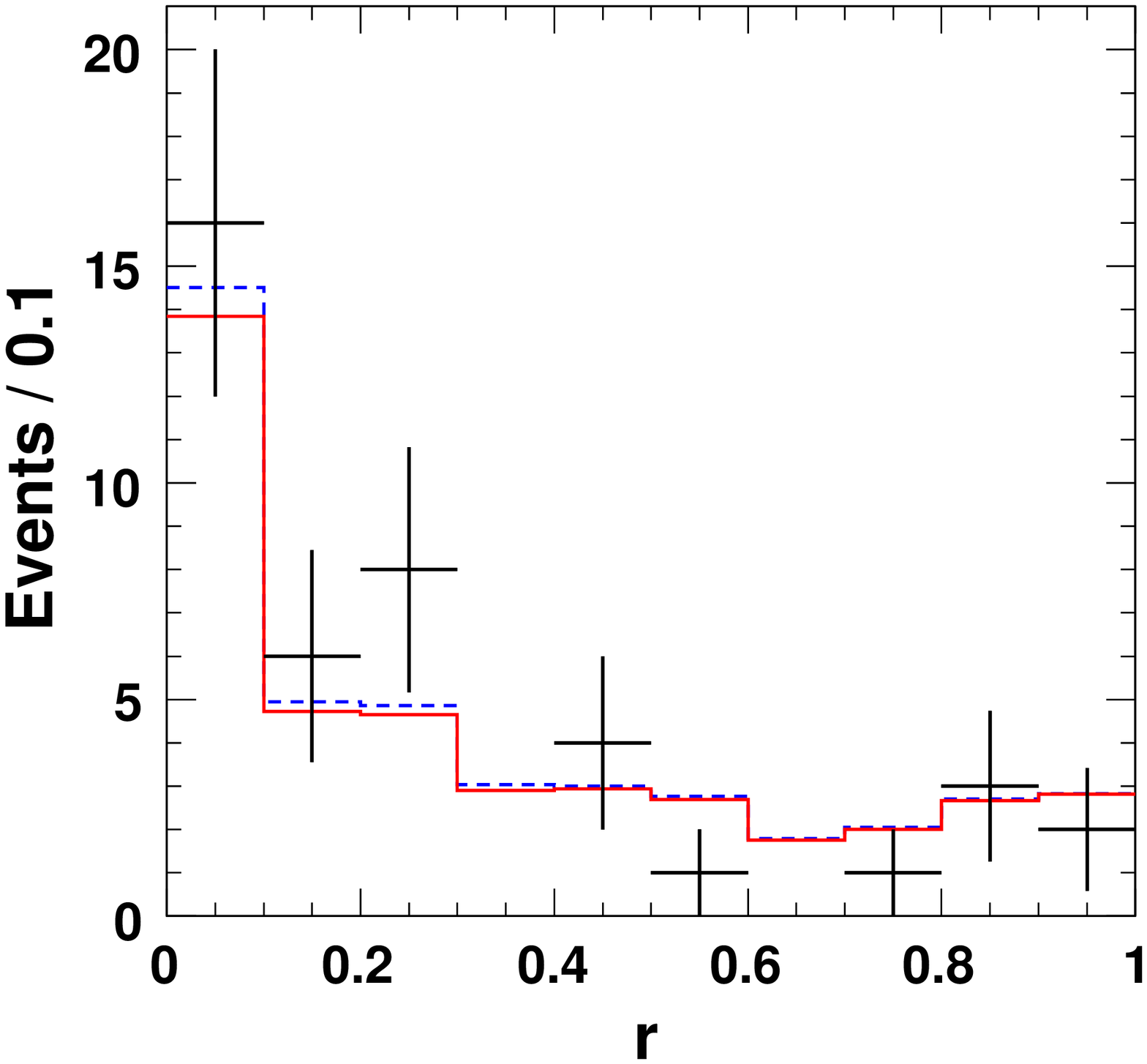}
\caption{
 Recoil mass (upper) and $r$ (lower) distribution 
 for samples reconstructed as
 $\Upsilon(4S) 
 \to 
 \left(
 J/\psi K^0_S ~,~ (J/\psi, \eta_c) K^0_S
 \right)
 $ decay.
 The solid lines show the fits to signal plus background
 distributions while the dashed lines show the background distributions.
}
 \label{fig_dat_jpsiks}
\end{figure}
The results of the final fit are shown in Fig.~\ref{fig_dat_jpsiks}.
The extracted signal yield, 
$-1.5~^{+3.6}_{-2.8}$
events,
is consistent with 
zero as well as with the SM prediction (1.7 events).
An upper limit is determined 
with a frequentist method~\cite{FeldmanCousins}, where
the PDFs are smeared to include systematic uncertainties.
We obtain 
$
 {\cal B}\left(
 \Upsilon(4S) 
 \to B^0\bar{B}^0
 \to 
 J/\psi K^0_S , (J/\psi, \eta_c) K^0_S
 \right)
~
<
~
4
\times 10^{-7} 
$
 at the 90~\% confidence level,
where the SM prediction is 
$
1.4
\times10^{-7}$.
This corresponds to $F < 2$ at the 90~\% confidence level.
We also search for 
$
\left(
J/\psi K^0_S,~
J/\psi K^0_S
\right)
$
combinations by fully reconstructing both $B$ mesons.
No candidates are observed.

In summary, a search for $CP$ violation in $\Upsilon(4S)$ decays
was performed.
In a data sample of 535 million $B\overline{B}$ pairs obtained 
via decays of the $\Upsilon(4S)$ resonance,
no significant signals were observed.
We obtain an upper limit of 
$
4
\times 10^{-7} 
$
 at the 90~\% confidence level
for 
the branching fraction of 
the $CP$ violating modes,
$
 \Upsilon(4S) 
 \to B^0\bar{B}^0
 \to 
 J/\psi K^0_S + (J/\psi, \eta_c) K^0_S
$.
Assuming the SM, 
with an integrated luminosity of 
30
ab$^{-1}$ 
that is expected to be available in a future $B$ factory,
these decays can be observed with 5$\sigma$ significance.

We thank the KEKB group for excellent operation of the
accelerator, the KEK cryogenics group for efficient solenoid
operations, and the KEK computer group and
the NII for valuable computing and Super-SINET network
support.  We acknowledge support from MEXT and JSPS (Japan);
ARC and DEST (Australia); NSFC and KIP of CAS (China); 
DST (India); MOEHRD, KOSEF and KRF (Korea); 
KBN (Poland); MES and RFAAE (Russia); ARRS (Slovenia); SNSF
(Switzerland); 
NSC and MOE (Taiwan); and DOE (USA).


\begin{thebibliography}{99}
\bibitem{CC}
Throughout this paper, 
the inclusion of the charge conjugate mode decay is implied
unless otherwise stated.
\end{thebibliography}
\end{document}